\documentclass[final]{cvpr}

\usepackage{array}
\usepackage{booktabs}
\usepackage{times}
\usepackage{epsfig}
\usepackage{graphicx}
\usepackage{amsmath}
\usepackage{amssymb}
\usepackage{nicefrac}
\usepackage{algorithmic}
\usepackage{textcomp}
\usepackage{multirow}
\usepackage{placeins}
\usepackage{cite}
\usepackage{soul}

\usepackage{url,lineno,microtype,subcaption,amsmath,ctable,breakurl,comment,float}

\usepackage[pagebackref=true,breaklinks=true,colorlinks,bookmarks=false]{hyperref}

\DeclareMathSymbol{:}{\mathord}{operators}{"3A}

\def\cvprPaperID{22} 
\def\confYear{CVPR 2021}

\begin{document}

\title{GAN-Based Data Augmentation and Anonymization for\\ Skin-Lesion Analysis: A Critical Review}

\author{Alceu Bissoto\textsuperscript{1} ~~~Eduardo Valle\textsuperscript{2} ~~~Sandra Avila\textsuperscript{1}\\
\textsuperscript{1}Institute of Computing (IC) ~~~\textsuperscript{2}School of Electrical and Computing Engineering (FEEC)\\ 
RECOD Lab., University of Campinas (UNICAMP), Brazil\\
}

\maketitle

\begin{abstract}
Despite the growing availability of high-quality public datasets, the lack of training samples is still one of the main challenges of deep-learning for skin lesion analysis. Generative Adversarial Networks (GANs) appear as an enticing alternative to alleviate the issue, by synthesizing samples indistinguishable from real images, with a plethora of works employing them for medical applications. Nevertheless, carefully designed experiments for skin-lesion diagnosis with GAN-based data augmentation show favorable results only on out-of-distribution test sets. For GAN-based data anonymization — where the synthetic images replace the real ones — favorable results also only appear for out-of-distribution test sets. Because of the costs and risks associated with GAN usage, those results suggest caution in their adoption for medical applications.
\end{abstract}

\section{Introduction}
\begin{table*}[tbh]
\centering
\scriptsize
\begin{tabular}{p{0.005cm}l>{\raggedright}p{1.75cm}>{\raggedright}p{2.125cm}l>{\raggedright}p{2.8cm}ll>{\raggedright\arraybackslash}p{1.45cm}}
\toprule
  & Ref$_{\mathrm{Year}}$  &  GAN  &  Target Model  &  Domain  &  Dataset  &  Real:Synth.  &  Sampling  &  Gain/Metric\dag  \\
\midrule
  \parbox[t]{1mm}{\multirow{16}{*}{\rotatebox[origin=c]{90}{Classification}}} & \cite{bissoto2018skin}$_{2018}$  &  pix2pixHD, PGAN  &  Inception-v4  &  skin lesions  &  ISIC 2018 Tasks 1–2  &  1:1, 1:2  &  random  &  1 / AUC  \\
  & \cite{frid2018gan}$_{2018}$  &  DCGAN  &  custom CNN  &  liver CT  &  private (182 samples)  &  1:0.6  &  random  &  7.1 / acc  \\
  & \cite{wu2018conditional}$_{2018}$  &  translation-based  &  ResNet-50  &  mammography  &  DDSM  &  1:0.111 to 1:1  &  fixed  &  0.9 / AUC  \\
  & \cite{bisla2019towards}$_{2019}$  &  DCGAN  &  ResNet-50  &  skin lesions  &  ISIC 2017  &  1:0.26  &  random  &  4 / AUC \\
  & \cite{bozorgtabar2019informative}$_{2019}$  &  translation-based  &  VGG-16, ResNet-18, DenseNet  &  chest x-ray  &  ChestXRay14  &  learned  &  Bayesian NN  &  —  \\
  & \cite{han2019combining}$_{2019}$  &  PGAN, MUNIT, SimGAN  &  ResNet-50  &  brain MRI  &  BraTS 2016  &  1:0.96  &  —  &  1.7 / acc  \\
  & \cite{de2020assisting}$_{2020}$  &  DCGAN  &  LeNet-5, AlexNet  &  esophagus  &  MICCAI 2016 EndoVis Challenge  &  1:1, 1:5  &  random  &  2 / acc  \\
  & \cite{ghorbani2020dermgan}$_{2020}$  &  pix2pix-based  &  MobileNet  &  skin lesions  &  private (49920 clinical images)  &  1:0.4  &  balancing  & -11.3–13.4 / F1-score  \\ 
  & \cite{levine2020synthesis}$_{2020}$  &  PGAN  &  VGG19  &  histopathology  &  TCGA, OVCARE  &  1:22.8  &  random  &  4 / balanced acc \\
  & \cite{ye2020synthetic}$_{2020}$  &  —  &  ResNet-34  &  histopathology  &  PatchCamelyon, private (2647 patches)  &  RL-based  &  RL-based  &  4.6 / AUC  \\\cmidrule(rl){1-9}
  \parbox[t]{1mm}{\multirow{12}{*}{\rotatebox[origin=c]{90}{Segmentation}}} & \cite{shin2018medical}$_{2018}$  &  pix2pix  &  GAN-based  &  brain MRI  &  BraTS 2015  &  1:1*  &  fixed  &  1 / Dice  \\
  & \cite{bailo2019red}$_{2019}$  &  pix2pixHD  &  FCN-8s  &  red blood cells  &  private (100 images)  & 1:1*  &  fixed  &  0.3 / Dice  \\
  & \cite{sandfort2019data}$_{2019}$  &  CycleGAN  &  U-Net  &  abdomen CT  &  Kidney NIH, (Liver, Spleen) DataDecathlon  &  1:0.06 to 1:2.6  &  random  &  1–60 / ood Dice   \\
  & \cite{amirrajab2020xcat}$_{2020}$  &  translation-based  &  modified U-Net  &  cardiac MRI  &  ACDC, SCD, York Cardiac MRI, private test set  &  1:0.66  &  fixed  &  1–11 / Dice\\
  & \cite{li2020tumorgan}$_{2020}$  &  translation-based  &  cascaded net, U-Net, DeepLab-v3  &  brain MRI  &  BraTS 2017  &  1:1  &  fixed  &  3 / Dice \\
  & \cite{mahapatra2020pathological}$_{2020}$  &  translation-based  &  U-Net, DenseUNet  &  opt. coher. tomogr.  &  RETOUCH  &  0:1*  &  fixed  &  2 / Dice  \\
  & \cite{pollastri2020augmenting}$_{2020}$  &  DCGAN, LAPGAN  &  custom CNN  &  skin lesions  &  ISIC 2017  &  —  &  random  &  2 / Jaccard  \\
  & \cite{jiang2020covid}$_{2021}$  &  translation-based  &  U-Net  &  covid lung CT  &  Radiopaedia  &  1:0.1 to 1:.0.5  &  fixed  &  2.5 / Dice \\ 
\bottomrule
\end{tabular}
\caption{The 18 works selected for our analysis of existing art. —: unclear or missing. *: missing details. \dag{}All gains in percentage points over the metric, min–max: ranges of gains found in experiments, ood:out of distribution, acc: accuracy.}, 
\label{tab:literature}
\end{table*}

The lack of training images is perhaps the main challenge faced by medical deep learning, deep skin-analysis being no exception. Although the availability of high-quality public datasets has mitigated the issue~\cite{mendoncca2013ph, isic2018data, isic2019data, isic2020data, ballerini2013color, Kawahara2018-7pt}, the total number of annotated skin-lesion images available to researchers is still 1–2 orders of magnitude smaller than the size of general-purpose computer vision datasets \cite{ILSVRC15,sun2017revisiting}.

More training images translate to better results, especially for deeper network architectures~\cite{valle2018data, menegola2017transfer}. However, annotating skin-lesion images to increase datasets process is very costly, depending on the scarce time of medical specialists. Generative Adversarial Networks (GANs) \cite{goodfellow2014} appear as an alternative to increase the amount of training data without incurring those costs.

GANs aim to artificially synthesize samples that are indistinguishable from real images. They may be employed as a complement to traditional data augmentation~\cite{perez2018data}, artificially increasing the amount of training samples. A plethora of existing works, which we discuss in our literature survey, suggest applying them for that purpose, but — as we will show — obtaining reliable improvements from GAN-based data augmentation is \textit{far} from obvious: in our carefully designed experiments, GAN-based augmentation failed to \textit{reliably} improve the classification performance of skin-lesion diagnosis, although we obtained good results for selected special cases. 

Another potential application of GANs is data anonymization, where the synthetic images are used to \textit{replace}, instead of augment, the original training set. Here again, our results recommend caution, although they were promising for out-of-distribution tests.

The main contribution of this work is a detailed study of the factors that can impact GAN-based augmentation, including GAN architectures, amount of real images used, proportion of real and synthetic images, and method for sampling synthetic images. The procedure adopted, from GAN checkpoint selection until classification network evaluation, can serve as a guideline to increase the reliability of future works using GAN-based augmentation. In addition, we carry out a systematic literature review, where we summarize the techniques used for GAN-based augmentation. In that review, we list issues in experimental design that may lead to overoptimistic results. 

The text follows the usual organization, with the literature review next, proposed approach in Section~\ref{sec:method}, followed by results in Section~\ref{sec:experiments}. We close the paper with a discussion of the main findings, the risks of using GAN-based augmentation in medical applications, as well as cautious avenues for continuing their use in this context.

\section{Literature review} \label{sec:relatedwork}

The review in this section started to seed an attempt to obtain reliable performance improvements from GAN-based data augmentation. Since several existing works reported measurable gains, we explored literature to understand which factors previous authors had tested in their experiments. For more details over different GAN methods, we direct the reader to surveys in the area \cite{shamsolmoali2021image, wang2021generative}.

However, as our experiments progressed without revealing reliable improvements, we returned to literature to subsidize a larger-scale experiment. Our review grew, in scope and formality, proportionally to our experimental ambitions. Although this review does not intend to be a formal meta-analysis, we took inspiration from Preferred Reporting Items for Systematic Reviews and Meta-Analysis (PRISMA)~\cite{prisma2009} to gather a representative sample of existing art.

Our starting point was all GAN-related works published in the past ISIC Workshops. To that  we added a database search, in Google Scholar, with the query ``GAN generative\_adversarial\_networks medical\_image synthetic\_image data\_augmentation classification OR segmentation \text{-NLP} \text{-temporal} \text{-tabular}'', which gave 251 results. Notice that we did not restrict our query to skin-lesion analysis but to all medical-image applications. We excluded from the sample all works outside our scope (\textit{i.e.}, no GAN data augmentation, no test on a medical dataset, no classification or segmentation), as well as works without experimental results. We excluded surveys and reviews from our sample as well. 
To get a manageable sample of papers to study, we kept our sample only works from top conferences (CVPR, NeurIPS, MICCAI), and their respective workshops, or published in journals of impact factor 3.0 or higher. For the same reason, we did not include unpublished preprints.

The resulting collection of 18 works appears in Table~\ref{tab:literature}. The table details which GANs each work evaluated, which deep-learning model was the target of the data augmentation, what was the medical application domain, which datasets were used on the evaluation, and which improvements the authors reported over which metrics. 

Depending on the domain, dataset, and task at hand, different families of GANs may be more appropriate, or conversely, completely unfeasible. Translation-based GANs, which include pix2pix~\cite{pix2pix}, pix2pixHD~\cite{pix2pixhd}, and SPADE~\cite{park2019semantic}, 
learn to translate between different types of images, \textit{e.g.}, from a segmentation mask into a new synthesized input, or from a non-contrast to a contrast CT-scan.  Their main advantage is that adherence to the mask tends to improve their biological/medical coherence.

In contrast, noise-based generation models, like DCGAN~\cite{radford2015dcgan} and PGAN~\cite{karras2017progressive} offer flexibility, the latter being able to generate high-resolution images. Noise-based generation may suffer, however, if the training sets are too small, requiring mitigating techniques such as patch extractions and traditional data augmentation. 

\begin{table*}[tbh]
\small
\centering
\begin{tabular}{l>{\raggedright}p{2.75cm}>{\raggedright}p{8.75cm}>{\raggedright\arraybackslash}p{2.25cm}} 
\toprule
Application  &  GANs  &  Real:Synthetic Ratios\newline(Real:Synthetic Benign:Synthetic Malignant)  &  Sampling Technique  \\
\midrule
Augmentation  &  pix2pixHD, PGAN, SPADE, StyleGAN2  &  $1:0$, $1:\nicefrac{1}{4}$,  $1:\nicefrac{1}{2}$, $1:1$, $(1:\nicefrac{1}{2}:\nicefrac{1}{2})$, $(1:\nicefrac{1}{2}:\nicefrac{3}{4})$, $(1:\nicefrac{1}{2}:1)$  &  random, diverse, best, worst \\
Anonymization  &  StyleGAN2  &  $\nicefrac{1}{16}:0$, $\nicefrac{1}{16}:\nicefrac{1}{16}$, $\nicefrac{1}{16}:\nicefrac{15}{16}$, $\nicefrac{1}{8}:0$, $\nicefrac{1}{8}:\nicefrac{1}{8}$, $\nicefrac{1}{8}:\nicefrac{7}{8}$, $\nicefrac{1}{4}:0$, $\nicefrac{1}{4}:\nicefrac{1}{4}$, $\nicefrac{1}{4}:\nicefrac{3}{4}$, $\nicefrac{1}{2}:0$, $\nicefrac{1}{2}:\nicefrac{1}{2}$, $1:1$, $1:0$  &  random   \\
\bottomrule
\end{tabular}
\caption{Summary of the two proposed experiments, with the main factors and their levels.}
\label{tab:exps}
\end{table*}

During the review, we took freehand notes about the experimental protocol of each paper, to understand which factors literature considered the most important to evaluate, and which varied the most among works. We found the proportion of real training images to images synthesized by the GAN to be one of the foremost factors, as well as the technique used to sample the images from the GAN.

In addition to requiring an input to serve as a guide for generation, translation-based GANs tend to have a limited range of outputs, each input being able to create a single synthetic image (which is why most of them are marked as \textit{fixed} for sampling in Table~\ref{tab:literature}). For that same reason, works using them tend to employ a $1:1$ proportion of real:synthesized images, although $1:<1$ ratios are also possible. 
Works using noise-based GANs tend to display more diverse choices of sampling and image ratios. 

In our freehand comments, we also noticed possible issues with experimental protocols that we wished to avoid in our large-scale experiments. The main issues we found amidst existing art were: (1) giving to the GAN-augmented models better data-access than to the baseline model, especially by choosing hyperparameters directly on the test set; (2) (hyper)optimizing the GAN-augmented models more thoroughly than the baseline models; (3) failing to use best current training practices on the baseline model, \textit{e.g.} best available (conventional) data-augmentation, learning-rate choice, normalization, \textit{etc}.; (4) ignoring performance fluctuations, \textit{e.g.}, by performing a single run, or by failing to report the deviation statistics. Not all works suffer from all those issues, of course, but we believe they may explain the discrepancy between the results we report next, and those found in current art.

We decided to limit the scope of this review to works that apply GANs to medical applications, and not to the GANs themselves. We remark, however, that choosing the GAN model for medical-image augmentation puts stringent requirements on the model. First, the model must be able to generate high-resolution images, to accommodate the visual patterns that characterize medical images. For skin-lesion analyses, the patterns that differentiate benign or malignant skin lesions are rather fine-grained, and state-of-the-art classification networks have inputs from $224\times224$ to $1024\times1024$ pixels. Second, the model must able to generate class-conditional samples, \textit{i.e.}, to create synthetic samples which convincingly belong for each of the dataset classes, so those may join the supervised training dataset coherently.

\section{Experiment design} \label{sec:method}

\begin{table*}[tbh]
\small
    \centering
    \begin{tabular}{lrl>{\raggedright}p{5.5cm}>{\raggedright\arraybackslash}p{4.5cm}}
        \toprule
         Test Dataset & Size & Imaging Tech. & Diagnostic Classes & Notes  \\
         \midrule
         isic19~\cite{isic2019data} & $3,863$ & Dermoscopic & melanocytic nevus, melanoma, benign keratosis, actinic keratosis, dermatofibroma, vascular lesion & in-distribution, same classes as train \vspace{.05cm}\\
         isic20~\cite{isic2020data} & $1,743$ & Dermoscopic & melanocytic nevus, melanoma, benign keratosis, actinic keratosis, lentigo, benign unknown & mainly in-distribution, potential out-of-distribution samples among the ‘benign unknown’ samples\vspace{.05cm}\\
         derm7pt–derm~\cite{Kawahara2018-7pt} & $872$ & Dermoscopic & melanocytic nevus, melanoma, seborrhoeic keratosis & out-of-distribution, fewer classes than train \vspace{.05cm}\\
         derm7pt–clinic~\cite{Kawahara2018-7pt} & $839$ & Clinical & melanocytic nevus, melanoma, seborrhoeic keratosis & out-of-distribution, fewer classes than train \vspace{.05cm}\\
         dermofit~\cite{ballerini2013color} & $973$ & Clinical &  melanocytic nevus, melanoma,  seborrhoeic keratosis, actinic keratosis,  pyogenic  granuloma, haemangioma, dermatofibroma & out-of-distribution benign classes \vspace{.05cm}\\
         \bottomrule
    \end{tabular}
    \caption{Description of the test sets used in the evaluation of the classification networks trained with the augmented training~set.}
    \label{tab:testsets}
\end{table*}

When we started this study, the initial goal was to maximize the performance of GAN data-augmentation for skin-lesion analysis. As our preliminary experiments progressed, we found our results to be extremely noisy. Performance improvement, when it happened at all, was completely random: the choice of GAN model or other factors had no explanatory power. We changed our research question to a more fundamental one: \textit{can GAN data-augmentation actually improve the performance of skin-lesion analysis?} Failing that, \textit{can GANs be used to anonymize the training data?} The latter application — using synthetic samples \textit{instead of} the actual real data — would be profitable even with a small, tolerable drop in accuracy, since it could make feasible for different countries and institutions to share knowledge in situations where direct patient images could not be exchanged (\textit{e.g.}, due to incompatible privacy laws). We design two experiments (Table \ref{tab:exps}) to answer those questions, which we detail~below.

\paragraph{GAN-based data augmentation.} As seen in the previous section, literature shows no consensus on how to perform GAN-based data augmentation. Our experimental design attempted to reflect the diversity of approaches found in existing art, contemplating a diversity of GANs, real:synthetic image ratios, real training dataset sizes, and synthetic sampling techniques. We present those choices next.

\begin{figure*}[tbh]
\centering
\includegraphics[width=0.195\linewidth]{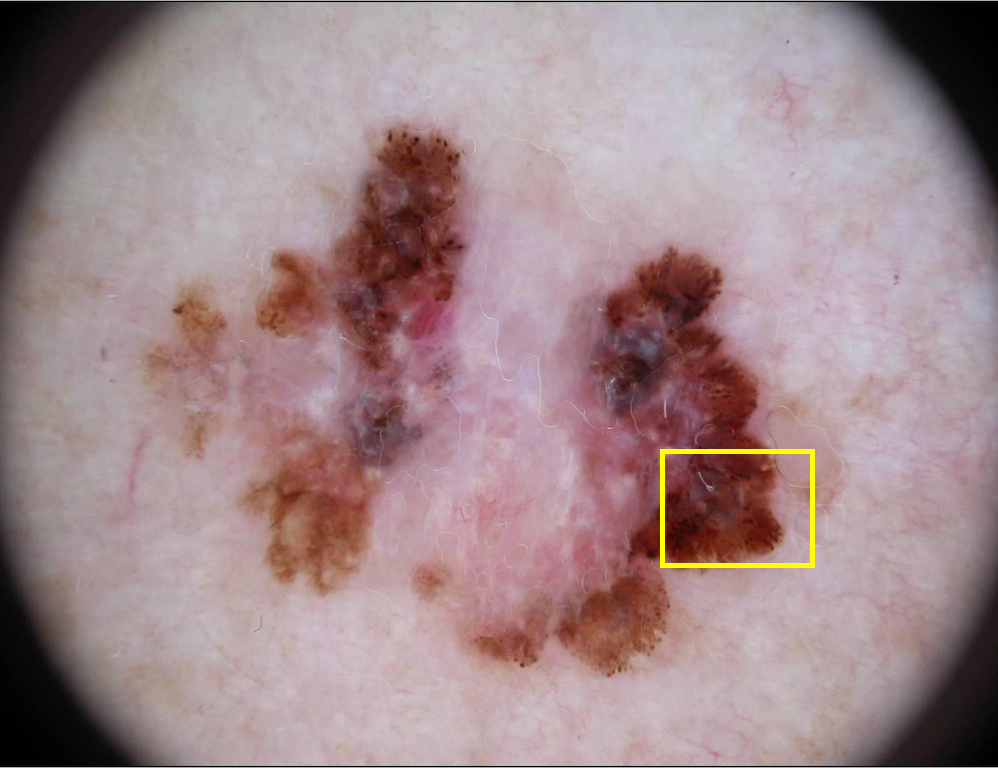}
\includegraphics[width=0.195\linewidth]{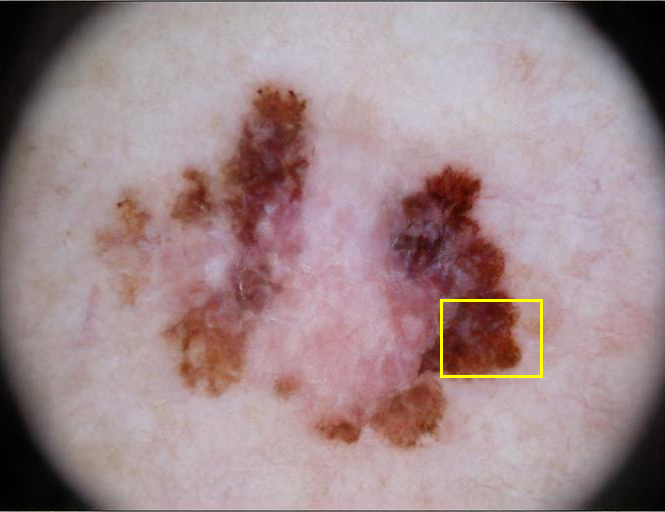}
\includegraphics[width=0.195\linewidth]{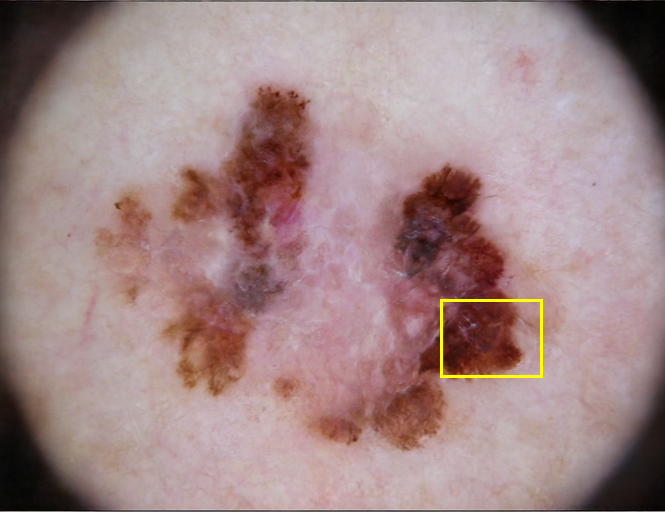}
\includegraphics[width=0.195\linewidth]{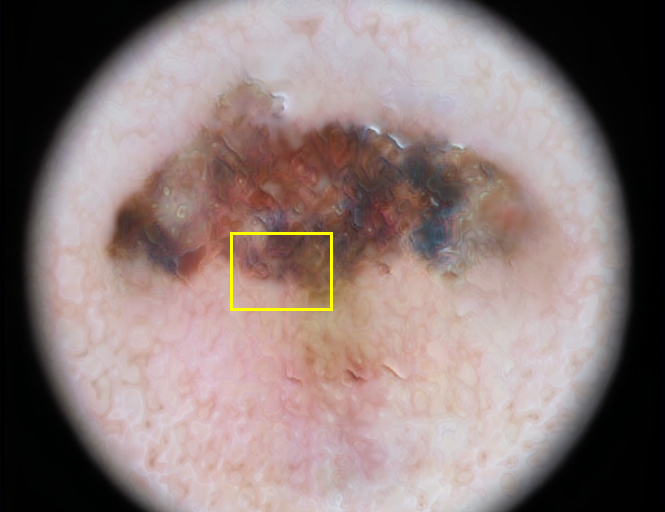}
\includegraphics[width=0.195\linewidth]{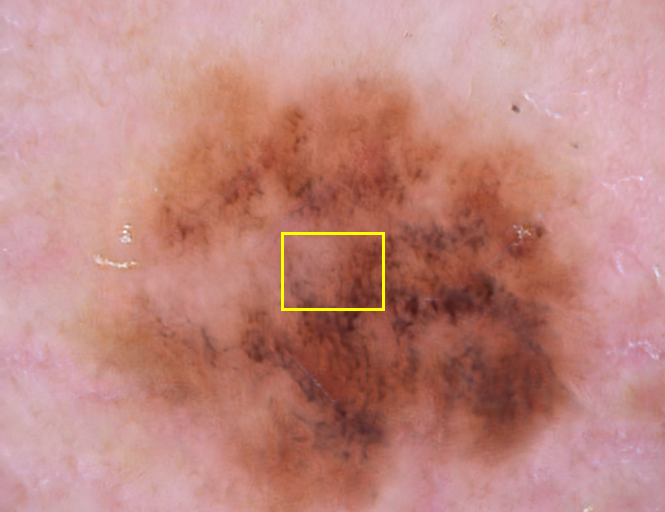}

\subcaptionbox{Real}{ \hspace{-0.17cm}
\includegraphics[width=0.195\linewidth]{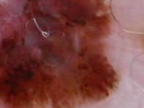}
} \hspace{-0.27cm}
\subcaptionbox{pix2pixHD}{
\includegraphics[width=0.195\linewidth,trim=1px 0 0 0,clip]{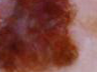}
} \hspace{-0.27cm}
\subcaptionbox{SPADE}{
\includegraphics[width=0.195\linewidth,trim=1px 0 0 0,clip]{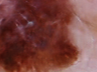}
} \hspace{-0.27cm}
\subcaptionbox{PGAN}{
\includegraphics[width=0.195\linewidth,trim=1px 0 0 0,clip]{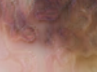}
} \hspace{-0.27cm}
\subcaptionbox{SytleGAN2}{
\includegraphics[width=0.195\linewidth]{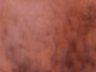}
}

\caption[Synthetic samples for different GAN-based approaches.]{Synthetic samples  for different GAN-based approaches: (a) Real, (b) pix2pixHD, (c) SPADE, (d) PGAN, \text{(e)~StyleGAN2}. In the first row, we present the full image, while in the second, we zoom-in to focus on the details.}\label{gans}
\end{figure*}

The GAN models investigated were pix2pixHD~\cite{pix2pixhd}, PGAN~\cite{karras2017progressive}, SPADE~\cite{park2019semantic}, and StyleGAN2~\cite{karras2020style}. We chose pix2pixHD and PGAN because they are known to work on skin-lesion data augmentation \cite{bissoto2018skin}, while SPADE and StyleGAN2 are considered the state of the art on image generation. Pix2pixHD and SPADE are translation-based, while PGAN and StyleGAN2 are noise-based techniques. While the former tend to generate very high-quality images, they have stringent limitations on the amount of images they can generate, due to the requirement of using segmentation masks or different image modalities as inputs. Noise-based techniques have no such limitations, but tend to generate images with lower visual quality, and risk reproducing artifacts (\textit{e.g.}, vignettes, rulers) that may reinforce biases in the dataset~\cite{bissoto2019constructing}.
For pix2pixHD and SPADE, we use the whole training set's mask to generate our synthetic set. For PGAN and StyleGAN2, we sample enough images to keep a $1:\nicefrac{1}{2}$ ratio. In Figure~\ref{gans}, we show lesions generated by the GANs used in our investigation. For pix2pixHD and SPADE, we use a mask from the training set as input, causing the synthetic images to be almost identical to the real ones for the human eye (although features from deep networks may be still considerably different between the two kinds of images).

For real:synthetic ratios, we considered the ratios $1:\nicefrac{1}{4}$, $1:\nicefrac{1}{2}$, and $1:1$, where $1$ is the size of our original real-image training set ($14,805$ samples). We considered additional experiments varying the proportion of benign and malignant synthetic images proportion, to evaluate the opportunity of using GAN-based data augmentation to correct class-imbalance~\cite{ghorbani2020dermgan}. Those experiments are notated (real:synthetic benign:synthetic malignant), and we considered the ratios  $(1:\nicefrac{1}{2}:\nicefrac{1}{2})$ (which is the same as $1:\nicefrac{1}{2}$ and has no balancing effect), $(1:\nicefrac{1}{2}:\nicefrac{3}{4})$, and $(1:\nicefrac{1}{2}:1)$. The baseline for all the experiments in this group is the ratio $1:0$, \textit{i.e.}, the experiment with the entire real training set and no synthetic data augmentation.  

We varied the amount of \textit{real} images employed in training mainly as an evaluation of GAN-based anonymization (see below). Those experiments may also be interpreted as the impact of GAN-based data-augmentation for different training-set sizes. We used real-image training set sizes with fractions of $\nicefrac{1}{16}$, $\nicefrac{1}{8}$, $\nicefrac{1}{4}$, $\nicefrac{1}{2}$, and $\nicefrac{1}{1}$ (the baseline) of the whole dataset.

Finally, we investigate different ways of sampling the synthetic images from the generated pool. That only applies to the noise-based GANs (PGAN and StyleGAN2), which can generate a limitless amount of images. We generated $100,000$ images for each of the $2$ classes (benign and melanoma), and evaluated different methods to select the ones to compose the training set: choosing them at \textit{random}, choosing them at random but with a criterion of \textit{diversity} inspired on the perceptual-sensitive hash (pHash) to exclude near-duplicates, or choosing the ones \textit{best}-classified (lowest error) or \textit{worst}-classified (highest error) by an ancillary skin-lesion classification model.

\paragraph{GAN-based anonymization.} In anonymization, instead of using GANs to augment the training set, synthetic images \textit{replace} real images. That application has received less attention in literature~\cite{mahapatra2020pathological}, but could be invaluable for researchers and institutions wishing to share knowledge while having to navigate issues of patient confidentiality.

For that experimental design, we evaluate a single GAN (StyleGAN2) and a single sampling technique (random), explained above. In contrast, we evaluate many more real:synthetic ratios in this experiment, varying the amount of both kinds of images, in order to evaluate the situation where an institution was training a model with its own real images, adding synthetic images from a GAN provided by another institution. For each group of experiments with a fraction of the initial real training-set $\nicefrac{1}{x}$, we evaluated experiments with ratios $\nicefrac{1}{x}:0$ (the baseline for the group, with no synthetic images), $\nicefrac{1}{x}:\nicefrac{1}{x}$ (doubling the training set), and $\nicefrac{1}{x}:\nicefrac{x-1}{x}$ (topping up the training set). We included the  $1:0$ ratio as a reference for the expected upper bound on the accuracy for those experiments.

\subsection{Datasets}

For all experiments, the reference training set of real images was based on the training set of the ISIC 2019 challenge~\cite{isic2018data, isic2019data}. We split that dataset into a training set ($14,805$ samples) and a validation set ($1,931$ samples) used in all our experiments, and a test set, with $3,863$ samples, added to our collection of test sets (\textit{isic19} on Table~\ref{tab:testsets}).

We trained the noise-based PGAN and StyleGAN2 with the entire training set, but pix2pixHD and SPADE require semantic segmentation masks to guide the generation. We employ the clinical attribute semantic masks of the ISIC Challenge 2018 Task~2, which are available for only $2,594$ images, or about $\nicefrac{1}{6}$ of the training set.

When fine-tuning the target models, fractions of real images refer to selections of that same training dataset (\textit{e.g.}, $\nicefrac{1}{4}:0$ would refer to a training set of $3,701$ randomly selected samples from the $14,805$ real ones). Fractions of the synthetic dataset refer to the same \textit{size} of the real dataset, for a selection on the synthetic generated images (\textit{e.g.}, $0:\nicefrac{1}{4}$ would refer to a training set of $3,701$ randomly selected while keeping the real dataset class ratio, sampled from the $200,000$ synthetic images, half benign, and half malignant).

We perform our tests in five gold-standard datasets (Table~\ref{tab:testsets}), selecting the classes to always perform a melanoma-\textit{vs.}-benign task (carcinomas, if present in the dataset, are discarded from both training and testing). Having an array of test sets, both similar to our training set (“in-distribution”) and very different (“out-of-distribution”) is an attempt to mitigate the effect of dataset bias~\cite{bissoto2020debiasing,bissoto2019constructing,geirhos2020shortcut} and measure the models' generalization ability.
For derm7pt, we remove near duplicates, and keep only classes present in the ISIC 2018 Challenge Task 2 dataset (melanoma, sebhorreic keratosis, and nevus). Those alterations result in $872$ samples for derm7pt-derm and $839$ samples for derm7pt-clinic. For dermofit, we remove the carcinomas, leaving $973$ samples.

\paragraph{Analyses.} We replicated each experiment ten times, varying the selection of the real images from the training set and repeating the fine-tuning of the target model. In all experiments, the metric was the area under the ROC-curve of the target model (AUC) for melanoma-\textit{vs.}-benign classification. A visual analysis of the results is given by blotting the individual data-points superimposed with a box-plot that, as usual, reveals the medians, quartiles, and range. We also plot the arithmetic means (red dots). There are two sets of plots, one for each application (augmentation and anonymization), plots were separated per test dataset, and within each plot, experiments were grouped (blue and black colors) to facilitate comparison. The labels of the experiments reveal the proportion of real:synthetic — or (real:synthetic benign:synthetic malignant) — images used in the training set used to fine-tune the target model, with additional information for the choice of GAN (p2p: pix2pixHD, spd: SPADE, pgn: PGAN2, sgn: StyleGAN2, all: samples from all GANs together), and the choice of sampling method (wst: worst, bst: best, div: diverse). When omitted, the GAN is StyleGAN2 and the sampling method is random.

\subsection{Implementation details}

For PGAN and pix2pixHD, we follow Bissoto \emph{et al.}’s \cite{bissoto2018skin} implementation, modifications, and hyperparameters. We also adopt their pix2pixHD generation procedure to SPADE. For StyleGAN2, we do not need to adapt the original GAN model implementation\footnote{\url{https://github.com/NVlabs/stylegan2}}, as it could generate class-conditioned samples from the start.

Evaluating the quality of synthetic images is an open problem. Available metrics often fail to evaluate diversity, and both broad- and fine-grained details. Most popular metrics (\textit{e.g.}, FID~\cite{heusel2017gans}, Inception Score~\cite{salimans2016}) rely on ImageNet-trained networks to extract characteristics, and are, thus, questionable for applications where classes are very different, such as medical applications. We address that issue using a broad analysis, comprising traditional GAN metrics, metrics on classifier models, and visualization techniques of the trained classifier.

To select the best training checkpoint for the GAN model, we considered both the time spent on training, and the FID metric~\cite{heusel2017gans} preferring later checkpoints (longer training) for similar FID ($<0.3$ difference). The exact training lengths, and corresponding FIDs, appear in Table~\ref{fid}.

The target model for both the data-augmentation and anonymization experiments is a skin-lesion classification Inception-v4~\cite{szegedy2016inceptionv4}, pre-trained on ImageNet. We chose this model based on its well-known performance on skin-lesion analysis, including its ability to profit from larger training sets~\cite{valle2018data}. The full set of weights of the target model is fine-tuned with stochastic gradient descent with momentum of $0.9$, weight decay of $0.001$ and learning rate of $0.001$, which we reduce by $10$ on each validation plateau after a $10$-epoch patience until a minimum of $10^{-5}$. We use a batch size of $32$, shuffling the data before each epoch.  We resize input images to $299\times299$ pixels, 
and z-normalize the input with ImageNet’s training set mean and standard deviation. We train for a maximum of 100 epochs, and an early stopping with a patience of 22 epochs.
We apply (conventional) data augmentation to \textit{all} experiments both during train \textit{and test}, with random horizontal and vertical flips, resized crops containing $75–100\%$ of the original image area,  rotations from -$45–45^{\circ}$ degrees, and hue changes of $-10–10\%$. We apply the same augmentations on both train and test. For test augmentation we average the predictions of $50$ augmented replicas of each sample~\cite{perez2018data}. We run each experiment 10~times, and in each one, we vary the real images, but keep the synthetics the same (following the sampling criteria).

Full-reproducible source code is available in our repository \url{https://github.com/alceubissoto/gan-aug-analysis}.

\begin{table}
\begin{center}
\begin{tabular}{lcc}
\toprule
GAN Architecture  & ~Epochs & ~FID \vspace{0.05cm}\\
\toprule
SPADE & 300 & 16.62 \\
pix2pixHD & 400 & 19.27 \\
PGAN & 890 & 39.57 \\
StyleGAN2 & 565 & 15.98 \\
\bottomrule
\end{tabular}
\caption{Amount of epochs and FIDs for each of the generative models used in this research. To select the checkpoint used to generate the samples that compose our classification model's training dataset, we consider both FID, and time spent training the GAN.}\label{fid}
\end{center}
\end{table} 
\section{Results} \label{sec:experiments}

In this section, we evaluate GAN-based augmentation procedures for skin lesion analysis.
In all the following experiments, we want to make the comparisons as fair as possible, giving equal opportunity to all models to be at their peak performance. 

\subsection{Augmentation}

The experiments for GAN-based data augmentation appear on the top row of Figure~\ref{fig:resultaug}. Comments on how to read and interpret those plots appear in Section~\ref{sec:method}, under the heading “Analyses”. 

The leftmost experiment in each plot, with label $1:0$, is the baseline with no synthetic-data augmentation. The plots reveal that, for dermoscopic test images, augmented-train sets are unable to confer a significant advantage, with augmented models showing lower — or at best similar — to the baseline. The experiments suggest that the more synthetic images we add to the training set, the worse the results are. The performance of different GANs fluctuated across datasets, but translation-based GANs tended to work better than noise-based GANS — but please notice that this factor is slightly confounded with image proportion in our tests due to the very limited generation ability of translation-based GANs. Of the two noise-based GANs, StyleGAN2 performed better (or at least, less worse).  

The scenario was less clear for the derm7pt-clinic dataset, where most experiments significantly improved the results. Those results departed from the other datasets also by showing StyleGAN2 ahead of all other GANs. However, the results on the dermofit dataset, also with clinical images, were more similar to the results on the dermoscopic datasets than to the ones on derm7pt-clinic.

We remind that the experiments on anonymization (explained next), may also be interpreted as experiments on data-augmentation for small training datasets, \textit{i.e.}, an anonymization experiment with ratio $\nicefrac{1}{x}:y$ can be reinterpreted as a data augmentation experiment with ratio $1:\nicefrac{x}{y}$ for a test dataset with $\nicefrac{1}{x}$ of the samples of our main one. As we will show, those experiments failed to reveal the advantage of synthetic augmentation, even for small datasets.

\subsection{Anonymization}

\begin{figure*}
    \centering
    \includegraphics[height=1.15\textwidth]{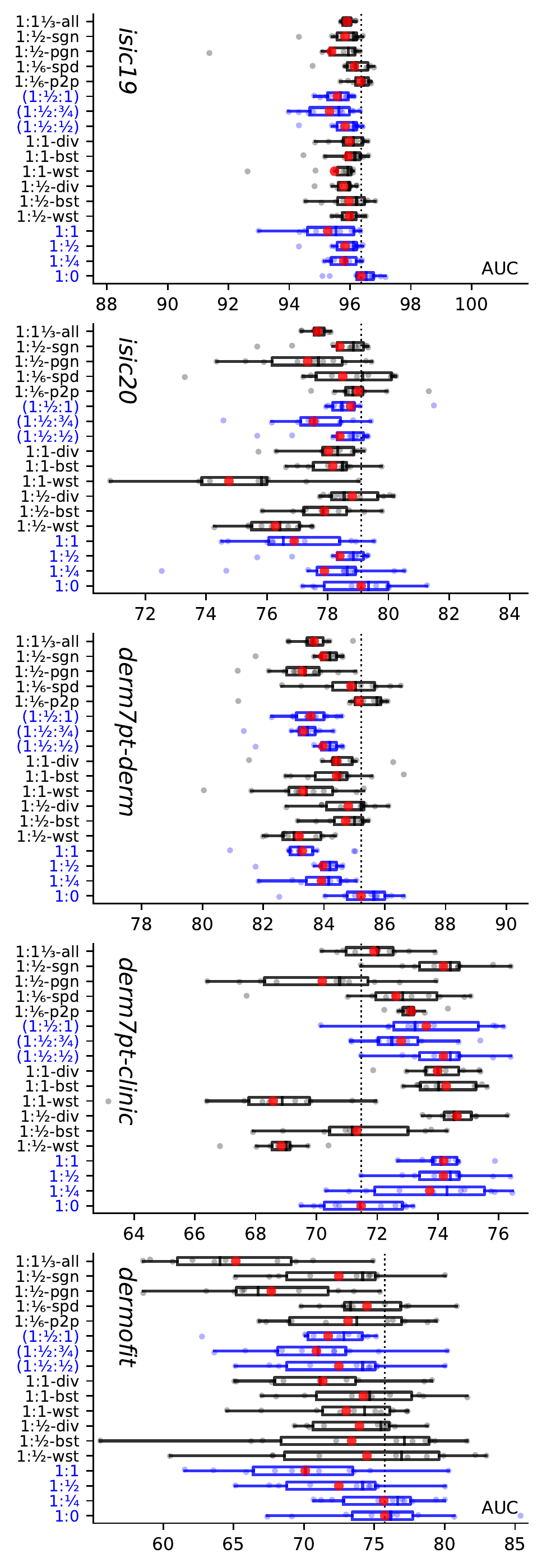}
    \includegraphics[height=1.15\textwidth]{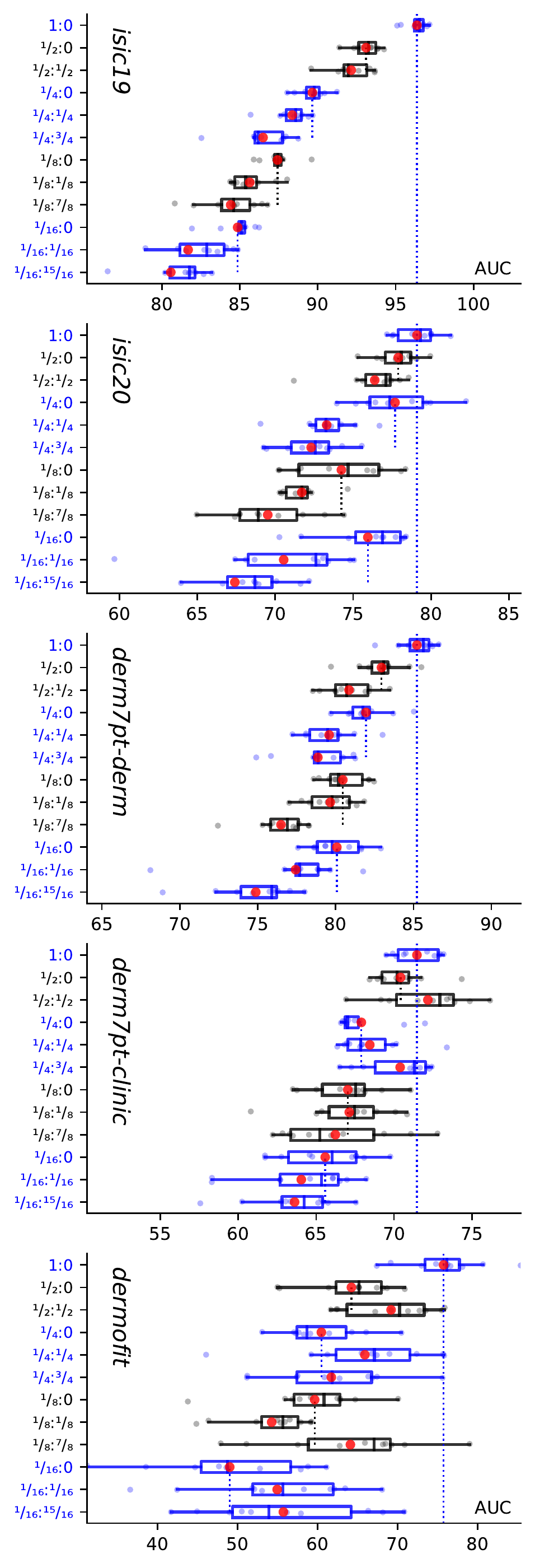}
    \caption{Results for GAN-based data augmentation (left) and GAN-based anonymization (right), separated by test dataset. Box-plots display medians, quartiles, and range, as usual; red dots show arithmetic means. Y-axis labels show real:synthetic or (real:synthetic benign:synthetic malignant) training image proportion, with additional information for the choice of GAN (p2p: pix2pixHD, spd: SPADE, pgn: PGAN2, sgn: StyleGAN2, all: samples from all GANs together), and the choice of sampling method (wst: worst, bst: best, div: diverse). When omitted, the GAN is StyleGAN2 and the sampling method is random. Experiments showcasing a factor are grouped together in alternating blue/black color.}
    \label{fig:resultaug}
\end{figure*}

The experiments for GAN-based anonymization appear on the bottom row of Figure~\ref{fig:resultaug}. Comments on how to read and interpret those plots appear in Section~\ref{sec:method}, under the heading “Analyses”. 

Those experiments are organized in groups of alternating blue and black colors, the rightmost experiment in each group, with label $\nicefrac{1}{x}:0$, being the baseline for the group. GAN-based anonymization only has interest if the performance of the experiment is significantly above that baseline — otherwise, it can be trivially obtained by simply discarding the problematic samples. The rightmost experiment shows the model trained with all real samples, and gives an expected upper bound of accuracy for the plot.

Unfortunately, for dermoscopic test images, no experiment appeared above that baseline. The derm7pt-clinic was again an exception, showing three experiments above the baseline, and one even slightly above the expected upper bound. In this set of experiments, the results on the (also clinical-image) dermofit dataset were also positive, with several configurations above the baseline.

\section{Discussion} \label{sec:conclusion} 

Our results suggest future authors interested in GAN-based data augmentation to be conservative about expected results, and cautious about evaluation protocols.

GAN-based augmentation is a technique extensively explored in the literature for mitigating the scarcity of training images, being particularly interesting for the medical images community.
However, we found our preliminary experiments to be excessively noisy, and noticed flaws in some experimental protocols during our literature review. Reliably transforming synthetic images into reliable performance gains is far from obvious.

We will \textit{not} go as far as \textit{condemning} GAN-based data augmentation. Our experiments suggest that for some specific out-of-distribution scenarios (\textit{e.g.}, training on isic19 and testing on derm7pt-clinic) the technique may provide reliable improvements. Characterizing exactly which scenarios are those is, however, still an open question, as experiments on the also clinical-image dermofit dataset did not confirm those findings. 

Because training GANs requires a huge computation investment — a single training of StyleGAN2 takes up to a month of GPU time — researchers and practitioners should carefully evaluate whether their application to data-augmentation is justifiable, considering, among other things, their energetic footprint~\cite{yu2019attributing}. The finicky nature of GAN training also brings other risks: missing modes~\cite{arjovsky2017wasserstein} may aggravate dataset biases, reinforcing the disparity of over/underrepresented groups instead of correcting them. GANs may also fixate on artifacts (such as vignettes, rulers, gel bubbles, \textit{etc.}) and introduce or reinforce spurious correlations on the data~\cite{bissoto2019constructing}. 

Our results for GAN-based anonymization show modest results, but here, at least, there seems to be a trend, with results for out-of-distribution data being generally favorable, and results for in-distribution data being generally unfavorable. Those data suggest that using GANs may be possible at least as an ancillary method for sharing knowledge while preserving patient privacy. For making that application safe, however, we need further studies on how much the GAN “remembers” each original training sample and on its ability to (purposefully or accidentally) reconstructing original samples.

Possible avenues for unblocking GAN-based data augmentation point towards attempting to conciliate the advantages of translation-based and noise-based techniques — obtaining the high-quality of the former and the limitless sampling availability of the latter — but such conciliation is a hard open problem. More achievable may be enhanced sampling methods, able to select the highest-quality, or better yet most relevant for decision, samples from the limitless sample of a noise-based technique. Even if future works fail to improve the ability of GANs to reliably provide  data-augmentation or anonymization, a better characterization of the cases they are able to improve may provide interesting insights on the fundamental workings of deep learning.

\section*{Acknowledgements}

A. Bissoto is partially funded by FAPESP 2019/19619-7. S. Avila is partially funded by CNPq PQ-2 315231/2020-3, and FAPESP 2013/08293-7. A.~Bissoto and S.~Avila are also partially funded by Google LARA~2020. E. Valle is funded by CNPq 315168/2020-0. This~project is partially funded by a CNPq Universal grant 424958/2016-3. This study was financed in part by the Coordenação de Aperfeiçoamento de Pessoal de Nível Superior -- Brasil (CAPES) -- Finance Code 001. \text{RECOD}~Lab. is supported by projects from FAPESP, CNPq, and~CAPES. We acknowledge the donation of GPUs by NVIDIA.

{\small
\bibliographystyle{ieee_fullname}
\bibliography{egbib}
}

\end{document}